\documentclass[twocolumn,prb,showpacs,preprintnumbers,amsmath,amssymb,superscriptaddress]{revtex4}%

\usepackage{graphicx}
\usepackage[usenames]{color}

\begin{document}

\title{Radiative cascade from quantum dot metastable spin-blockaded biexciton}
\author{Y.~Kodriano}
\affiliation{Department of physics, The Technion - Israel institute
of technology, Haifa, 32000, Israel}
\author{E.~Poem}
\affiliation{Department of physics, The Technion - Israel institute
of technology, Haifa, 32000, Israel}
\author{N.~H.~Lindner}
\affiliation{Institute of Quantum Information, Caltech, Pasadena, CA, 91125, USA}
\author{C.~Tradonsky}
\affiliation{Department of physics, The Technion - Israel institute
of technology, Haifa, 32000, Israel}
\author{B.~D.~Gerardot}
\affiliation{School of Engineering and
Physical Sciences, Heriot-Watt University, Edinburgh EH14 4AS,
United Kingdom }
\author{P.~M.~Petroff}
\affiliation{Materials Department, University of California Santa
Barbara, CA, 93106, USA}
\author{J.~E.~Avron}
\affiliation{Department of physics, The Technion - Israel institute
of technology, Haifa, 32000, Israel}
\author{D.~Gershoni}
\affiliation{Department of physics, The Technion - Israel institute
of technology, Haifa, 32000, Israel}
\email{dg@physics.technion.ac.il}

\date{\today}

\begin{abstract}
We detect a novel radiative cascade from a neutral semiconductor
quantum dot. The cascade initiates from a metastable biexciton state
in which the holes form a spin-triplet configuration,
Pauli-blockaded from relaxation to the spin-singlet ground state.
The triplet biexciton has two photon-phonon-photon decay paths.
Unlike in the singlet-ground state biexciton radiative cascade, in
which the two photons are co-linearly polarized, in the triplet
biexciton cascade they are cross-linearly polarized. We measured the
two-photon polarization density matrix and show that the phonon emitted
when the intermediate exciton relaxes from excited to ground state, preserves the exciton's spin. 
The phonon, thus, does not carry with it any which-path information other than its energy.
Nevertheless, entanglement distillation by spectral filtering was
found to be rather ineffective for this cascade.
This deficiency results from the opposite sign of the 
anisotropic electron-hole exchange interaction in the excited exciton
relative to that in the ground exciton.
\end{abstract}


\maketitle

\section{Introduction}

A quantum dot (QD) containing two electron-hole pairs - a biexciton,
returns to its vacuum state by emitting two photons in a radiative
cascade. The radiative cascade mostly discussed in the
literature\cite{Benson_PRL,Moreau,Regelman,Akopian06,MichlerNJP}
initiates from the (non-degenerate) ground state of the biexciton,
in which the two electrons and two holes form spin-singlets in their
respective lowest single carrier energy levels. The QD relaxes to
its vacuum state by spontaneously emitting two photons. The emission
of the first photon brings the system into the manifold of intermediate single
exciton states. Two out of the four possible single exciton states
are optically active (bright). If these two `bright exciton' states
were energetically degenerate, the two emitted photons would have been
polarization-entangled.\cite{Benson_PRL,MichlerNJP}. Usually, however, due to
the reduced symmetry of the QD, the bright exciton states are not
degenerate.\cite{Gammon96,Kulakovskii,Ivchenko_Pikus} This removes
the radiative cascade's `which-path' ambiguity. Consequently, the
polarization state of the emitted photon-pair contains mostly
classical correlations \cite{Moreau,Regelman}, and only a very
small, usually undetectable, degree of entanglement. The degree of
entanglement can be increased to a measurable level by spectrally
filtering out most of the unentangled photon pairs (those which are
only classically correlated), while keeping photon-pairs for which
which-path ambiguity exists.\cite{Akopian06,AkopianJAP,Meirom}

Here, we report on the observation of a new radiative cascade,
initiating from a \emph{metastable}, spin blockaded biexcitonic
state. The metastable biexciton state is composed of two electrons
in their ground state and two holes, one in its ground state and one
in an excited state. The two holes form a spin-triplet, and are
therefore blockaded from thermalization into their ground singlet
state by Pauli's exclusion principle. This metastable biexciton
radiatively decays to form a single exciton which contains a hole in
an excited state. The hole is no longer spin-blockaded and it
relaxes to its ground state, releasing its energy by emitting a
phonon. The ground-state exciton thus formed, spontaneously
recombines radiatively. Consequently, this cascade involves the
emission of two photons and one phonon. The intermediate
non-radiative decay is fast ($\sim$30 ps)\cite{Poem10}, and as we
show below, \emph{preserves the exciton's spin}.

For a spin-preserving non-radiative decay, one expects all the which-path 
information carried by the intermediate phonon to reside in its
energy. Therefore, one would naively expect spectral filtering to be effective in
restoring entanglement also in this cascade.

We show below that this is not the case. We applied the same
filtering scheme for the two types of radiative cascades, from the
same quantum dot, under the same conditions. However, whereas for
the ground-state, singlet biexciton cascade, entanglement was
restored, it was not restored for the triplet, spin-blockaded,
cascade. In Sec.~\ref{sec:5} below, we show that the opposite sign
in the fine-structure splitting of the two intermediate excitonic levels, combined
with the fluctuating electrostatic environment (`spectral
diffusion')\cite{Meirom}, is responsible for this deficiency.

\section{Energy levels}
\label{sec:2} 
\begin{figure}[tbh]
  \includegraphics[width=0.48\textwidth]{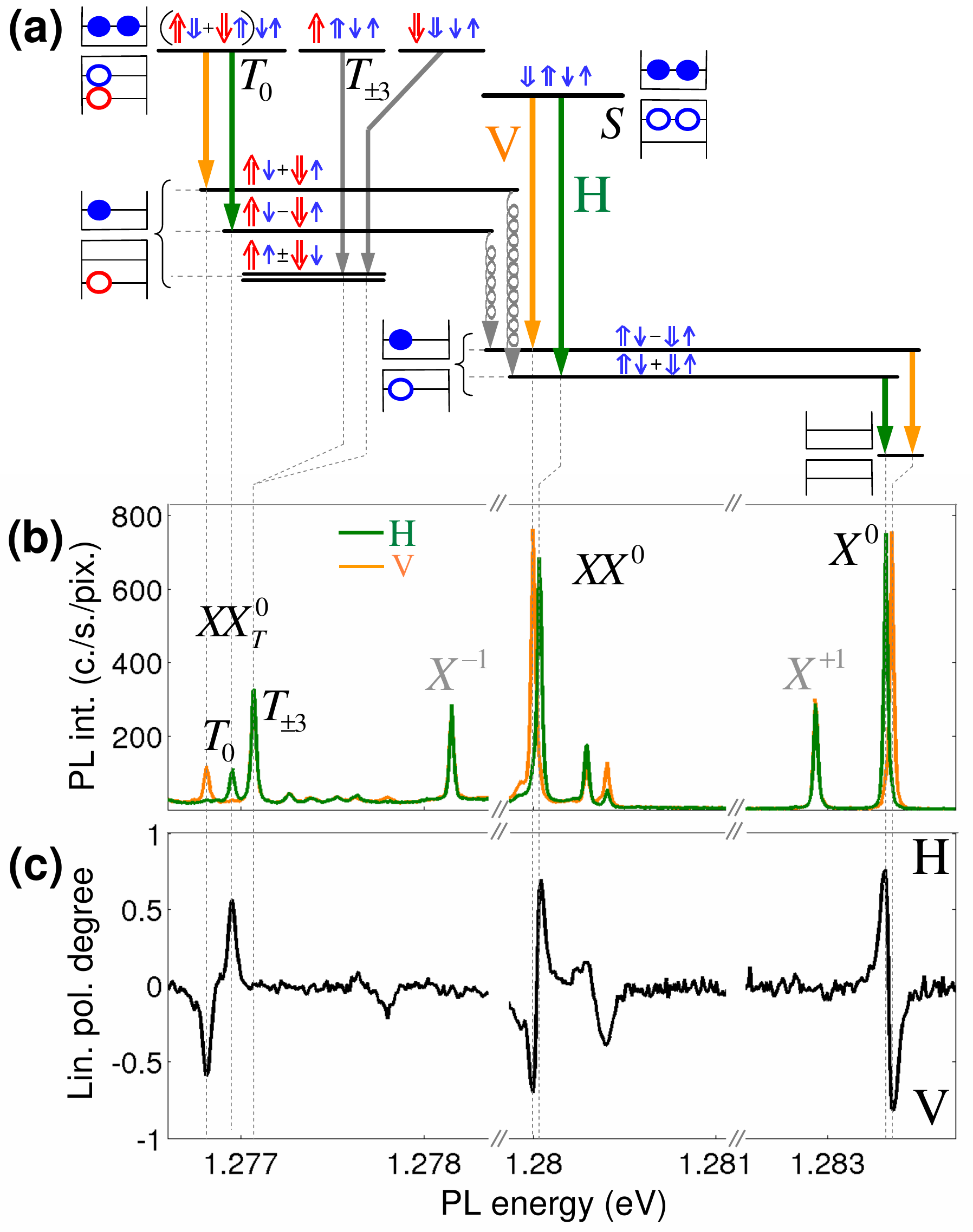}
  \caption{(color online) (a) Energy levels diagram for excitons and biexcitons in a
  neutral QD. Single-carrier level occupations are given along side each
  many-carrier level. The spin wavefunctions are depicted  above each level. The symbol
  $\uparrow$ ($\Downarrow$) represents spin up (down) electron (hole). Short blue
  (long red) symbols represent charge-carriers in the first (second) energy level.
  S indicates the ground biexciton state. T$_0$ (T$_{\pm3}$) indicates the
  metastable spin-triplet biexciton state with z-axis spin projection of 0
  ($\pm$3). The solid (curly) vertical arrows indicate spin preserving (non-)
  radiative transitions. Green (dark gray) [orange (light gray)] arrows represent
  photon emission in horizontal - H [vertical - V] polarization. (b) Polarized PL
  spectra. H (V) in green (orange). Spectral lines which are relevant to this work are marked
  and linked to the transitions in (a) by dashed lines.
  (c) Linear polarization spectrum. The value 1 (-1) means full  H (V) polarization.}
  \label{fig:1}
\end{figure}

A diagram of the relevant energy levels  of a neutral
QD is presented in Fig.~\ref{fig:1}(a). The ground-biexciton state
is marked by S (for Singlet). The metastable-biexciton states are
marked by T$_0$ and T$_{\pm3}$, where T stands for Triplet and the
subscripts stand for the total- two holes' spin projection on the
QD's growth direction. Since the electrons form a spin singlet in
their ground state, the electron-hole exchange interaction vanishes,
and the triplet states remain degenerate. Hole-hole anisotropic
exchange interaction is expected to remove this
degeneracy\cite{bimberg_anis_hhx} by lowering the T$_{\pm3}$ states
with respect to the T$_0$ state. This separation, in our case is
smaller than the isotropic electron-hole exchange interaction, which
removes the degeneracy between the dark and bright single exciton
states. Therefore, the energy order between the T$_0$ and T$_{\pm3}$
biexcitonic emission lines is the same as if there was no hole-hole
anisotropic exchange interaction, and we safely leave it out in the
following discussion.

Recombinations from the T$_{\pm3}$ states lead to the optically
inactive (`dark') exciton intermediate states [see the spectral line
in Fig.~\ref{fig:1}(b)]. Radiative decay does not proceed from these
states and they are discussed elsewhere\cite{Poemdarkexcitonoscilations}.

The ground state, singlet biexciton recombines through one of the
two single bright exciton states. Due to in-plane anisotropy, the
eigenstates of the bright exciton are the symmetric and
antisymmetric combinations of its $J=\pm 1$ spin-projection states
\cite{Ivchenko_Pikus} [see Fig.~\ref{fig:1}(a)]. Radiative
recombinations to these states and from them are thus co-linearly
polarized, as shown in Fig.~\ref{fig:1}(a), resulting in co-linear
polarization correlations between the cascading two photons
\cite{Moreau,Akopian06}.

Similarly, The T$_0$ biexciton state recombines to form an
{\em excited} bright exciton, in which the heavy hole resides in
its second energy level. The eigestates of this exciton are also the
symmetric and antisymmetric combinations of spin-projection states.
Here however, the energy order of the symmetric and antisymmetric
eigenstates is opposite to that of the ground exciton. The reason
for this order reversal is the difference between the spatial
symmetry of the ground and the excited hole wavefunction. The
anisotropic e-h exchange interaction, which induces this degeneracy
removal, is proportional to the quadrupole moment between the
electron and hole wavfunctions\cite{Gupalov,Takagahara,Poem07}. 
Therefore there is a sign difference
in the e-h exchange between the case in which the hole has s-like
symmetry (ground state exciton) and the case in which the hole has
p-like symmetry (excited exciton)~\cite{Gupalov,Poem07}.

There is yet another marked difference between the spin blockaded,
T$_0$ biexcitonic transition and the ground state, S one. The
difference is in the polarization selection rules for optical
transitions. In the ground state biexciton, the two holes form a
spin singlet, while in the T$_0$ one they form a triplet. Therefore,
while in the S-biexciton recombination, emission of horizontally (vertically) polarized photon
results in symmetric (anti-symmetric) bright exciton eigenstate, in the T$_0$-biexciton rcombination
it results in antisymmetric (symmetric) eigenstate of the excited
bright exciton.  A similar effect happens in the recombination of a
doubly-charged exciton\cite{Poem07,Warburton_doubly_charged}, there it is due to the final states.

Thus, the energy order of the biexciton doublet is the same for both
biexcitons. This results from the fact that the difference in
polarization selection rules is compensated by the sign difference
in the anisotropic exchange interaction. This is indeed what we
observe experimentally, as can be clearly seen in the polarized PL
spectra presented in Fig.~\ref{fig:1}(b), and in the corresponding
spectrum of linear polarization degree shown in Fig.~\ref{fig:1}(c).

The hole of the excited exciton states is not spin-blockaded and it
quickly relaxes (non-radiatively) to its ground state. If during
this relaxation, the exciton's spin state is preserved, horizontally
(H) polarized T$_0$ biexciton recombination will be followed by
vertically (V) polarized ground state exciton recombination (see
Fig.~\ref{fig:1}(a)). Thus, for the case of spin-preserving
non-radiative relaxation, the photon-pair emitted during the T$_0$
biexciton cascade should be \emph{cross}-linearly polarized, unlike
the co-linearly polarized photon-pair emitted in the S biexciton
cascade. Below we show that this is indeed the case.

\section{Experimental}
\label{sec:3} The studied sample was grown by molecular beam epitaxy
on a (001) oriented GaAs substrate. One layer of strain-induced
InGaAs QDs was deposited in the center of a 285 nm thick intrinsic
GaAs layer. The layer was placed between two distributed Bragg
reflecting mirrors, made of 25 (bottom) and 10 (top) periods of
pairs of AlAs/GaAs quarter wavelength thick layers. This constitutes
a one optical wavelength (in matter) microcavity for light emitted
due to recombination of QD confined e-h pairs in their lowest energy
levels. The microcavity increases light harvesting efficiency, for
emission which resonates with its cavity mode.

For the optical measurements the sample was placed inside a tube
immersed in liquid Helium, maintaining sample temperature of 4.2K. A
$\times 60$, 0.85 numerical aperture, in-situ microscope objective
was used both to focus the exciting beam on the sample surface and
to collect the emitted light. The collected light was split by a non
polarizing beam splitter, and each beam was dispersed by a 1~meter
monochromator and detected by either an electrically-cooled CCD
array detector or by a single channel, single photon, silicon
avalanche photodetector. The system provides spectral resolution of
about 10~$\mu eV$.

The polarization of the emitted light in each beam was analyzed
using two computer-controlled liquid-crystal variable retarders and
a linear polarizer. The retarders were carefully calibrated at the
emission wavelength such that cross talks between various
polarization projections never exceeded 5\%.

Standard photon counting electronics was then used to measure the
differences between the arrival times of two photons originating
from two different spectral lines, at given polarizations. The
histogram of these times, when normalized, gives the time-resolved
intensity correlation function. The response function of the system
and its temporal resolution ($\sim$0.4~ns) were determined by
measuring picosecond laser pulses~\cite{Regelman,Akopian06}.

\section{Results}
\label{sec:4} 
\begin{figure}[tbh]
  \includegraphics[width=0.48\textwidth]{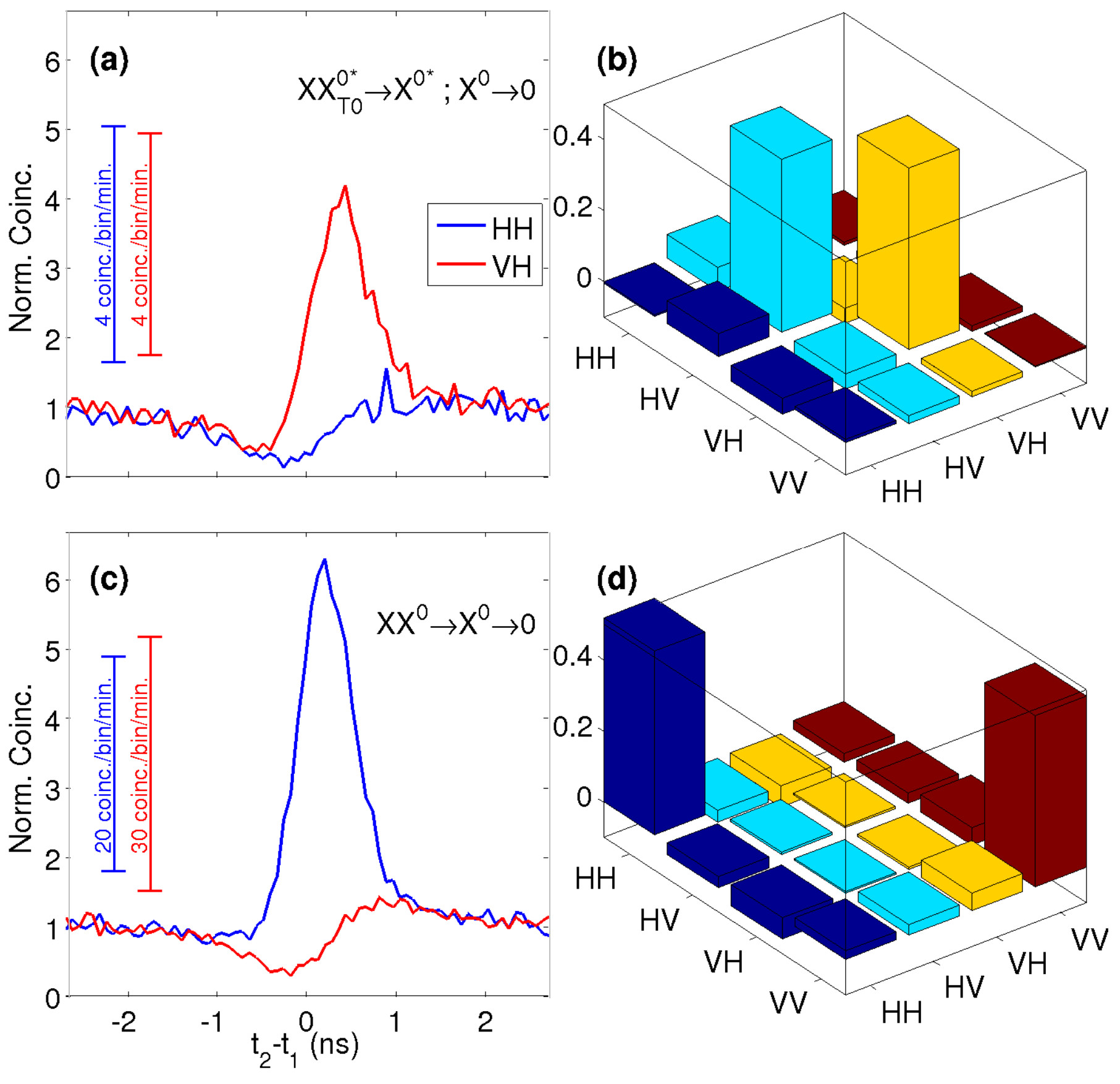}
  \caption{(color online) (a), (c) Measured intensity correlation functions for
  the spin-blockaded and ground-state biexciton cascades, respectively. Blue
  (dark gray) [red (light gray)] line represents the correlation in co- [cross-]
  linear polarizations. The coincidence rates are indicated by the scale-bars in
  units of coincidences per time bin (80 ps) per minute. (b), (d) Real parts of
  the two-photon polarization density matrices measured for the spin-blockaded
  and ground-state biexciton cascades, respectively.}
  \label{fig:2}
\end{figure}

In Fig.~\ref{fig:2}(a) [Fig.~\ref{fig:2}(c)] we
present measured time-resolved intensity correlation functions
between the T$_0$ [S] biexciton line and the exciton line, in both
co- (blue) and cross- (red) linear polarizations. It is clearly seen
that while for the cascade starting from the S biexciton, the
emission of the exciton is ``bunched'' for co-linear polarizations
and ``anti-bunched'' for cross-linear polarizations, the exact
opposite happens for the cascade starting from the T$_0$ biexciton.
This confirms our understanding of the T$_0$ biexciton cascade.
Moreover, our observations unambiguously demonstrate that there is
no change in the spin-state of the excited bright exciton during its
phonon assisted relaxation to the ground exciton state.

In addition, we performed full polarization tomography for both
cascades, with and without spectral
filtering.\cite{Akopian06,AkopianJAP,Meirom} The resulting two-photon
polarization density matrices for the case of no filtering are
presented in Fig.~\ref{fig:2}(b) [Fig.~\ref{fig:2}(d)] for the T$_0$
[S] biexciton cascade. Since the imaginary parts of the matrices
were zero to within the accuracy of the measurement, only the real
parts are displayed.

\begin{figure}[tbh]
  \includegraphics[width=0.48\textwidth]{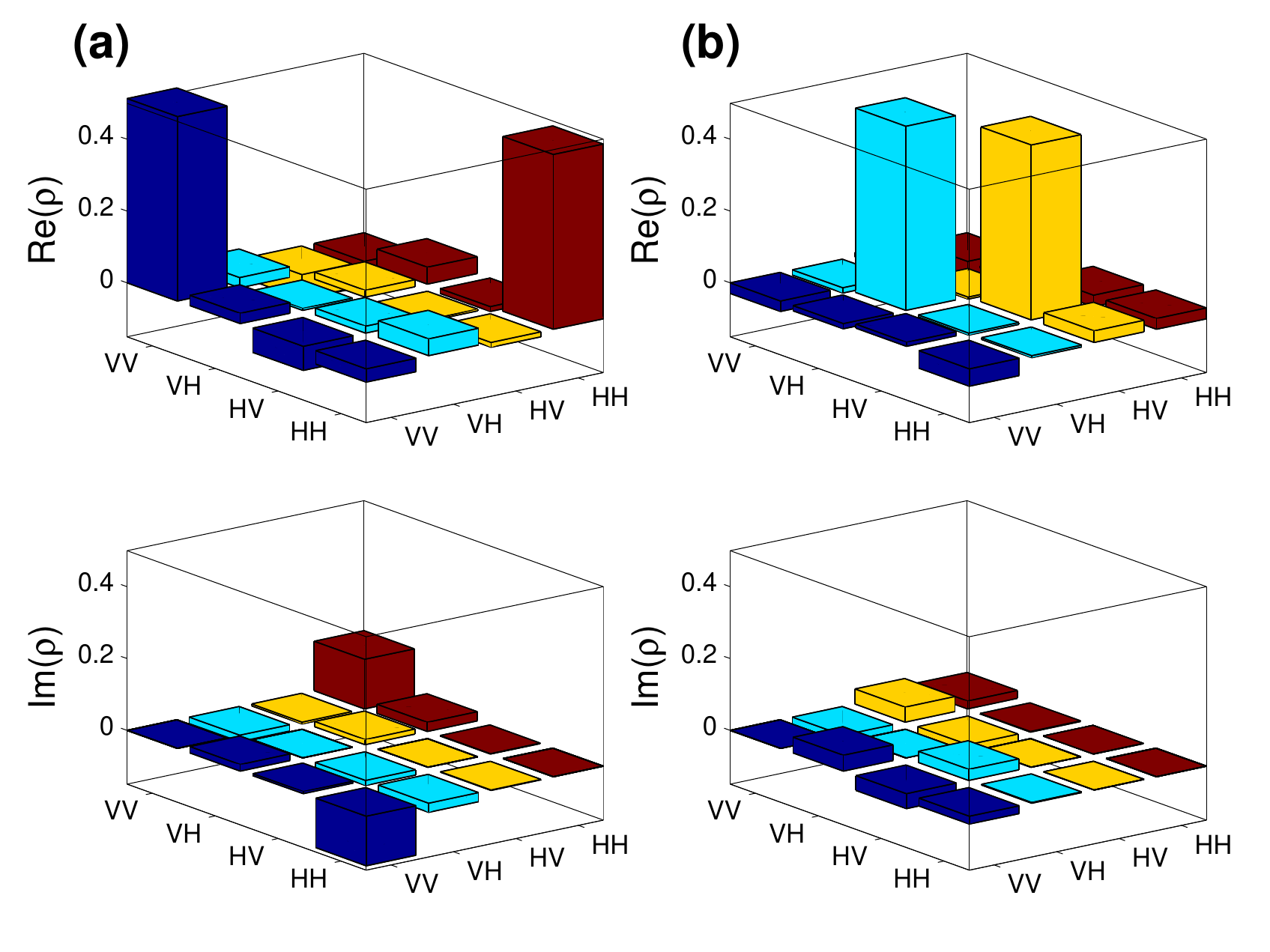}
  \caption{(a), (b) Measured two-photon polarization density matrices for
  the ground-state (S) and spin-blockaded ($T_0$) biexciton cascades,
  respectively, as obtained with spectral filtering. Real (imaginary) parts are shown
in the  top (bottom) panels. The Peres-criterion (negativity of the
partially transposed matrix) for the matrix in (a) [(b)] is
0.15$\pm$0.03 [0.05$\pm$0.1].}
  \label{fig:3}
\end{figure}

In Fig.~\ref{fig:3} we present the polarization density matrices
obtained for the case when spectral filtering is applied. While
entanglement is restored for photon pairs emitted from the S
biexciton cascade\cite{Akopian06} [Fig.~\ref{fig:3}(a)], no such
restoration is observed for the T$_0$ cascade [Fig.~\ref{fig:3}(b)].
The reason for this difference is not the emission of an additional
phonon in the latter cascade, as one would naively assume. The
emitted phonon does not interact with the relaxing heavy hole's spin, as our
data unambiguously demonstrate. Therefore the phonon does not carry
with it any 'which path' information (apart from its energy)\cite{Akopian06}, which would have rendered spectral filtering inefficient in restoring the entanglement. The
reason is more involved and we discuss it below.

\section{Discussion}
\label{sec:5}
\begin{figure}[tbh]
  \includegraphics[width=0.48\textwidth]{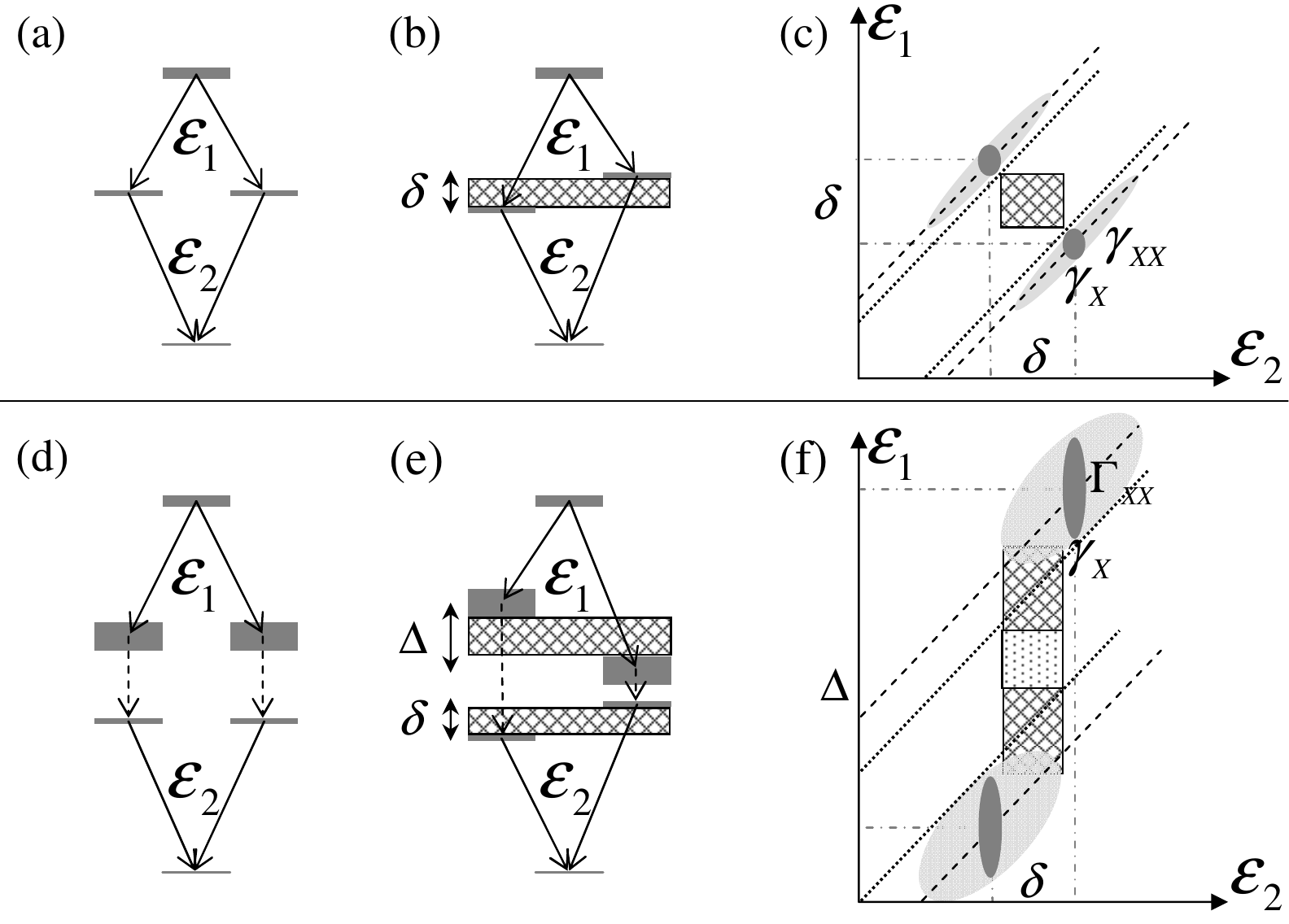}
  \caption{(a) Ideal direct cascade. The widths of the levels represent their
  decay rate. (b) Ground state biexciton cascade in an anisotropic QD. (c)
  Schematic two-photon probability distribution. Dark gray - high probability
  areas. Cross-hatched - spectral filter. Light gray - inhomogeneous broadening
  due to spectral diffusion. (d)-(f) Same as (a)-(c) (respectively), for an
  indirect cascade. In (e) and (f) only the case of splittings in opposite
  directions is shown. The dotted rectangle in (f) is an example for a filter
  not penetrated by the high-probability areas for any amount of spectral diffusion.}
  \label{fig:4}
\end{figure}

Fig.~\ref{fig:4}(a) shows the  cascade initiated by the ground state
of a biexciton in an ideal, symmetric QD. In this case, the two
exciton energy levels are degenerate, and emitted photon pairs will
be entangled.\cite{Benson_PRL,MichlerNJP}

Fig.~\ref{fig:4}(b) shows the case of the ground-state biexciton
cascade in an asymmetric QD, in which the exciton levels are split
by an energy $\delta$. Here, spectral filtering (cross-hatched
rectangle) is necessary for the emitted photons to be
entangled.\cite{Akopian06,AkopianJAP,Meirom}

Fig.~\ref{fig:4}(c) shows a schematic plot of the two-photon
probability distribution. The x- (y-) axis represents the energy of
the exciton (biexciton) photon. The dark-gray spots show the regions
of high emission probability. Their size and shape are determined by
the radiative width of the exciton ($\gamma_X$) and biexciton
($\gamma_{XX}$) lines. The emission in these regions is dominated by
un-entangled photon pairs.\cite{Akopian06,AkopianJAP,Meirom}
The energies of the two photons are
related by total energy conservation: if the first photon has high energy,
the second one will have low energy, and vice-versa. This puts the
two dark-gray spots on a line parallel to the (1,-1) direction. The
cross hatched rectangle represents an optimal spectral filter for
entanglement restoration. It is $\delta-\gamma_X$ by
$\delta-\gamma_{XX}$ in size. It rejects most of the un-entangled
photon-pairs while it keeps a measurable fraction of the entangled
pairs, which lie mostly between the two dark gray spots, on the
connecting line. The degree of entanglement within the filtered
photon pairs is thus increased.\cite{Akopian06,AkopianJAP,Meirom} A
narrower filter would yield higher degree of entanglement, but it
will transmit considerably less photon pairs.

Due to random fluctuations in the electrostatic environment of the
QD, the energies of the spectral lines fluctuate with time. This
``spectral diffusion'' happens on timescales much longer than the
radiative recombination time of the exciton. The random electric
field is thus quasi-static. Since all spectral lines experience
almost the same shift in a given static electric field,\cite{Ware}
the energies of the exciton and biexciton lines will fluctuate in a
correlated manner. The dark-gray spots of Fig~\ref{fig:4}(c) will
thus randomly move along the dashed lines parallel to the (1,1)
direction, as shown in the figure by the light-gray areas. As long
as these areas are strictly outside the filtered area, spectral
diffusion does not interfere with the entanglement restoration.
Indeed, as was demonstrated in ref.\cite{Akopian06} and here
[Fig.~\ref{fig:3}(a)], entanglement can be restored by spectral
filtering for the ground biexciton (S) cascade.

The situation is somewhat different for the spin blockaded cascade
reported here. Figures~\ref{fig:4}(d)-(f) describe this case.
Here there is a fast, non-radiative (phononic) relaxation of the
hole from its excited state to its ground state in between the
biexciton and exciton radiative recombinations. Fig.~\ref{fig:4}(d)
shows the ideal case where the excited and ground exciton states are
each two fold degenerate. The short lifetime of the excited exciton
states lead to a larger width ($\Gamma_{XX}$) of the corresponding
energy levels.

Since the spin of the exciton is conserved during the intermediate
stage, and since the spatial parts of the exciton wavefunctions
are identical for both decay paths, the emitted phonon does not
carry any `which path' information beyond its energy. In this case
one expects that appropriate filtering of the photon energies will
restore the which path ambiguity, resulting in entanglement of the
polarization state of the photons.

Fig.~\ref{fig:4}(e) shows the case of an asymmetric QD. The
degeneracy is lifted for both ground and excited exciton states,
albeit, as explained above, {\em in opposite directions}. Due to
the opposite sign of the degeneracy removal, in one path {\em both}
photons are more energetic, while in the other path both are less
energetic.

This is shown in Fig.~\ref{fig:4}(f), where the dark-gray spots
again represent the regions of high probability. Their elongated
shape is due to the larger width of the biexciton photon, which
comes from the fast decay of its final state. Spectral diffusion
will still shift these regions along the (1,1) direction, as shown
by light gray zones. The analog of the optimal filter for this case
is shown by the cross hatched rectangle. It is $\Delta-\Gamma_{XX}$
by $\delta-\gamma_X$, where $\Delta$ is the excited exciton
splitting (in absolute value). As in the previous case, such a
filter excludes the dark-gray spots. However, it is not immune to
spectral diffusion, as shown by the overlap of the cross hatched
rectangle and the light gray zones. Indeed, as shown in
Fig.~\ref{fig:3}(b), no entanglement could be detected even when
this filter was applied.

Further decreasing the filter width may solve the spectral diffusion
problem, as shown by the dotted rectangle in Fig.~\ref{fig:4}(f),
which completely avoids the light-gray zones and any other possible
location of the dark-gray spots. However, for such a filter the
photon pair collection rate would be drastically lower. We note that
if the ground and excited exciton states would have split to the
same direction, the situation would have been similar to that
described in Fig.~\ref{fig:4}(c), and entanglement restoration by
spectral filtering would not have been affected by spectral
diffusion, despite the phonon emission.

A quantitative condition for spectral filtering to effectively erase
the which path information can be formulated by inspecting
Figs.~\ref{fig:4}(c,f). Note first that spectral diffusion leads to
motion of the dark gray spots along the \mbox{(1,1)} direction.
Spectral filtering works if during this motion the spots stay
strictly outside the filter area. The width of the filter that
satisfies this condition is determined by looking at the projection
of the filter on the orthogonal direction to the motion of the
spots: the \mbox{(1,-1)} direction. Let $f>0\, (F>0)$ be the filter
width for the exciton (biexciton) photon spectral line. The filter's
projection on the (1,-1) direction is given by \mbox{$(f+F)/\sqrt
2$}. The projection of the line connecting the centers of the two
dark-gray spots on the (1,-1) direction is given by
\mbox{$|\Delta\pm\delta|/\sqrt 2$} where the plus sign is for
Fig.~\ref{fig:4}(c) and the minus sign is for Fig.~\ref{fig:4}(f).
For avoiding overlap one thus must have
\mbox{$F+f\leq|\Delta\pm\delta|$}. The case of a minus sign forces
narrow filter widths, which makes spectral filtering ineffective.
Taking into account the widths of the lines (the sizes of the
dark-gray spots), leads to a stronger constraint
    $$F+f\leq|\Delta\pm\delta|-(\Gamma_{XX}+\gamma_X)$$
where $\Gamma_{XX}$ refers also to $\gamma_{XX}$ as appropriate. In
this work, \mbox{$\delta=34\mu eV$}, \mbox{$\Delta=140\mu eV$},
\mbox{$\gamma_{X}=1.5\mu eV$}, \mbox{$\gamma_{XX}=3\mu eV$}, and
\mbox{$\Gamma_{XX}\approx40\mu eV$}. Therefore, for the singlet
biexciton cascade, where $\Delta=\delta$, the constraint becomes
\mbox{$F+f\leq2\delta-(\gamma_{XX}+\gamma_X)=63\mu eV$}, just
slightly smaller than \mbox{$2\delta=68\mu eV$}. For the triplet
biexciton cascade we obtain \mbox{$F+f\leq64\mu eV$}, but this is
now much smaller than \mbox{$\Delta+\delta=174\mu eV$}, and
therefore hardly yields any coincidences. The dotted rectangle in
Fig.~\ref{fig:4}(f) is an example of such a filter. For measuring
the density matrix in Fig.~\ref{fig:3}(b) we used
\mbox{$f\approx30\mu eV$} and \mbox{$F\approx100\mu eV$}. We could
not perform the experiment for a smaller $F$ due to diminishing
coincidences signal. This explains why while entanglement was
restored for the singlet biexciton cascade, it was not restored for
the triplet cascade.

\section{summary}
\label{sec:6} We identified optical transitions due to radiative
decays of metastable, spin blockaded hole-triplet biexciton states
in the PL spectrum of single neutral QDs. Photons emitted
specifically from the symmetric two holes triplet state with zero
spin projection ($T_0$) were found to be temporally correlated with
photons emitted from the ground exciton transition ($X^0$). In
contrast to the correlations between the ground-state biexciton (in
which the two holes form a spin singlet) and the exciton
transitions, `bunching' (cascaded emission) in this case was
observed for cross-linear polarizations. We explain this observation
in terms of the spin-exchange symmetry of the initial, biexciton
state: while the ground state biexciton is anti-symmetric under hole
spin exchange, the triplet biexciton is symmetric. This reverses the
polarization selection rules for the biexciton transitions.

The cascade involves an intermediate non-radiative relaxation. The
measured correlations: bunching (antibunching) for cross- (co-)
linear polarizations show that this phonon-mediated relaxation
preserves spin.

Two-photon polarization density matrices were measured for both
cascades, with and without spectral filtering.  Entanglement was
restored for the ground biexciton cascade when spectral filtering
was applied to it. Entanglement was not restored when spectral
filtering was applied to the spin-blockaded biexciton cascade. We
attribute this deficiency of spectral filtering to the reversed
energetic order of the excited exciton states. This reversed order
allows un-entangled photon-pairs to penetrate the spectral filters
during spectral diffusion of the emission lines.

\begin{acknowledgments}
The support of the US-Israel binational science foundation (BSF),
the Israeli science foundation (ISF), the ministry of science and
technology (MOST) and that of the Technion's RBNI are gratefully
acknowledged.
\end{acknowledgments}

\end{document}